\documentclass[doublecol,a4paper]{epl2} 

\title{Memory and rejuvenation in a spin glass}
\shorttitle{Memory and rejuvenation in a spin glass}

\author{R. Mathieu\inst{*}, M. Hudl, and P. Nordblad}
\shortauthor{R. Mathieu \etal}

\institute{
  \inst{*}  Electronic address: roland.mathieu@angstrom.uu.se\\
  \inst{} Department of Engineering Sciences, Uppsala University, Box 534, SE-751 21 Uppsala, Sweden 
}


\abstract{
The temperature dependence of the magnetisation of a Cu(Mn) spin glass ($T_g$ $\approx$ 57 K) has been investigated using weak probing magnetic fields ($H$ = 0.5 or 0 Oe) and specific thermal protocols. The behaviour of the zero-field cooled, thermoremanent and isothermal remanent magnetisation on (re-)cooling the system from a temperature (40 K) where the system has been aged is investigated. It is observed that the measured magnetisation is formed by two parts: (i) a temperature- and observation time-dependent thermally activated relaxational part governed by the age- and temperature-dependent response function and the (latest) field change made at a lower temperature, superposed on (ii) a weakly temperature-dependent frozen-in part. Interestingly we observe that the spin configuration that is imprinted during an elongated halt in the cooling, if it is accompanied by a field induced magnetisation, also includes a unidirectional excess magnetisation that is recovered on returning to the ageing temperature.
}

\begin{document}

\maketitle

\textbf{Introduction.} - Spin glasses continue to fascinate scientists. The origin and nature of some basic experimental observations, such as the ageing, memory and rejuvenation phenomena, are still being discussed and investigated\cite{ghost}. New experimental realisations of  model systems such as the three-dimensional ($3D$) XY and two-dimensional ($2D$) Ising model systems were recently discovered and characterised\cite{roland-esmo,roland-lsmo,nanopap}, and novel nano-structured mesoscopic spin glasses were investigated\cite{tetsuya}. Spin glasses and glassy properties were recently even associated with exchange-biased spintronic devices\cite{xb} and multiferroics\cite{kleemann}.

On the theoretical front, atomic and magnetic orderings of spin glasses were recently studied using first-principles calculations\cite{oleg}. Also, a new supercomputer with the ability to carry out Monte Carlo simulations up to experimental time scales\cite{parisi} was built, and may permit the theoretical observation of chaos and rejuvenation effects\cite{takayama}, and give insights on the validity of the different views on the nature of the spin-glass state\cite{bouchaud}.

The non-equilibrium properties of spin glasses can be investigated experimentally by recording the temperature-dependent magnetisation $M$ on reheating after spin configurations are imprinted while halting the cooling at constant temperatures $T_h$ below the spin-glass phase transition temperature $T_g$\cite{dcagmn,dcising}. These equilibrations, or ageings, are kept in memory on further cooling and retrieved on reheating. Due to the chaotic nature of the spin-glass phase, this memory of the equilibration at $T_h$ is observed only in a finite temperature range around $T_h$, defining ``memory dips'' with a finite width. Outside this temperature range, the magnetisation recovers its reference level and the system appears to be rejuvenated. These so-called dc-memory experiments have been employed e.g. to investigate spin-glass model systems\cite{dcagmn,dcising,ghost}, superspin glasses\cite{nanop}, geometrically frustrated systems\cite{eric-pyro} and exotic superconductors\cite{salee}.

It was shown recently that the dynamical properties of different glassy and superparamagnetic systems could be compared by employing specific in-field temperature cycling-protocols\cite{mamiya,gamnas}, in which the magnetisation is measured on reheating up to a given temperature after an initial cooling to the lowest temperature. $M$ is then measured on repeated cooling down and reheating the system from different increasing temperatures.\\

We have here combined the two above procedures, adding in-field temperature-cycling procedures to dc-memory experiments. Using a weak magnetic field (0.5 Oe) in the linear response regime to magnetise the sample we are able to distinguish between a part of the magnetisation that is controlled by the dynamic response of the spin glass and a part that is frozen in and fades away when heating above the temperature where it was attained. Our experiments surprisingly uncover a unidirectional excess magnetisation associated with the application of the magnetic field.\\

\textbf{Experimental.} - We here investigate the temperature-dependent magnetisation of a Heisenberg-like Cu(Mn) spin glass, recorded after specific protocols on a noncommercial low-field superconducting quantum interference device (SQUID)\cite{squids}. Small magnetic fields of $H$=0.5 Oe were employed to probe the magnetisation, yielding a linear response of the system\cite{orm}. The magnetic field is generated by a small superconductive magnet with time constant $\sim$ 1 ms, allowing the shortest observation time after field switch to be of the order of 0.2-0.3 s\cite{dcising,squids}. The heating and cooling rates were of the order of 1 K/min while measuring, while the initial cooling rate was about 5 K/min. The Cu(Mn) spin glass ($T_g$ $\sim$ 57 K) employed in the experiments was prepared by drop-synthesis method using an induction furnace.\\

\begin{figure*}[h]
\begin{center}
\includegraphics[width=0.46\textwidth]{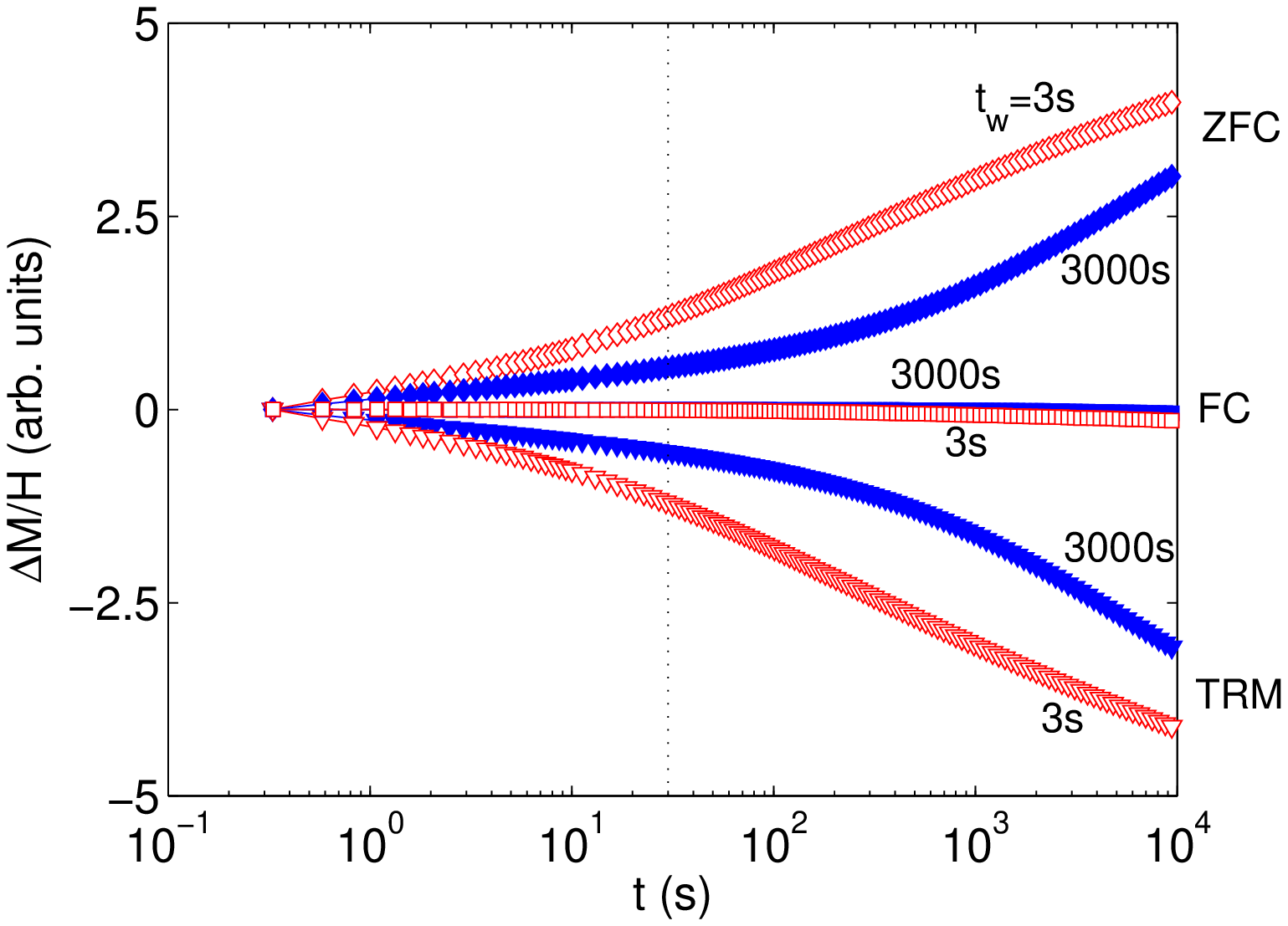}
\includegraphics[width=0.46\textwidth]{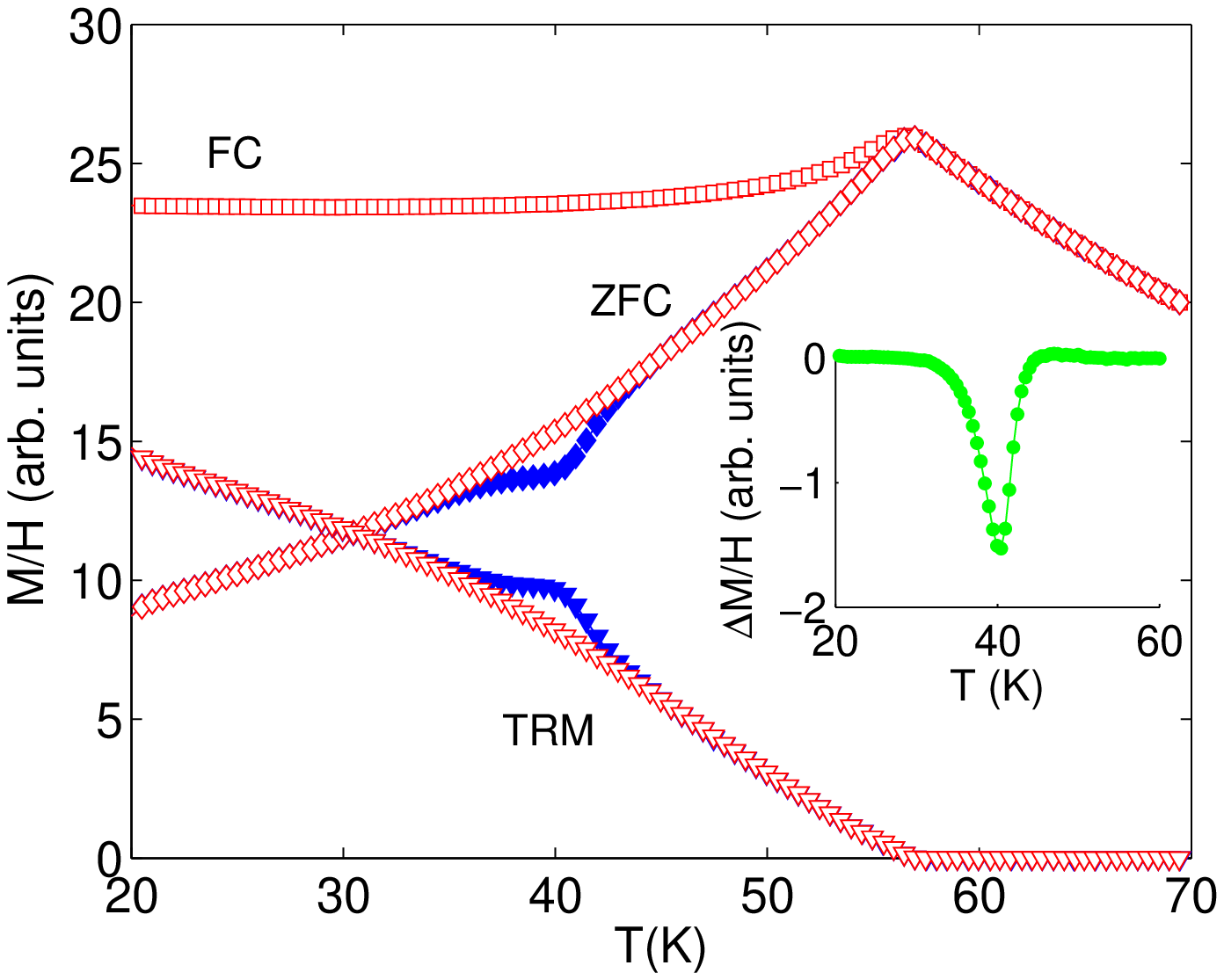}
\caption{Left: Time $t$ dependence of the ZFC, FC and TRM magnetisation $M$, recorded in $H$=0.5 Oe (ZFC, FC) or 0 Oe (TRM) after a quench from 70 K to 40 K, and a wait time $t_w$, plotted as $\Delta M=M-M(t=0.3 s)$. The vertical dotted line indicates the experimental time scale of temperature-dependent measurements. Right: Temperature $T$ dependence of the ZFC, FC, and TRM magnetisation. The $M(T)$ curves are measured on reheating after cooling the sample to the lowest temperature directly (open symbols, same symbols as in the left panel) or including a stop at 40 K for a time $t_s$=3000 s (filled symbols). A magnetic field $H$=0.5 Oe is employed to record the magnetisation. The inset shows the corresponding difference curve between ZFC measurements with and without stop at 40 K.}
\label{fig-intro}
\end{center}
\end{figure*}

\textbf{Results and discussion.} - The typical features of the ageing, memory and rejuvenation phenomena are exemplified in Fig.~\ref{fig-intro}. The left panel shows dc-relaxation experiments, in which the system is cooled down rapidly to a temperature below $T_g$, here 40 K. After a wait time of 3 s or 3000 s, a dc-field of $H$=0.5 Oe is applied and the magnetisation $M$ is recorded as a function of time $t$ in the case of zero-field cooled experiments. In the case of thermoremanent (TRM) experiments, the field is switched from 0.5 Oe to 0, while in the case of field-cooled (FC) ones the field is always 0.5 Oe. One can notice in the left panel of Fig.~\ref{fig-intro} that at an observation time of about 30 s, corresponding to the effective observation time of the magnetisation measurement on heating, the ZFC and TRM curves recorded after a 3000 s stop lies significantly below (resp. above) the curve recorded nearly immediately ($t_w$= 3 s) after reaching 40 K. This reflects the ageing or equilibration that occurred while the spin glass was left at a constant temperature, and the associated rearrangement of its spin configuration. The FC magnetisation $M_{\rm FC}$ shows in this context marginal relaxation behaviour, however, the principle of superposition is applicable and the fundamental relation: $M_{\rm ZFC}(t)$ $\sim$ $M_{\rm FC}(t)$ - $M_{\rm TRM}(t)$ is obeyed\cite{per}. The temperature (right)  and observation time (left) dependence of the different ZFC and TRM curves in Fig.~\ref{fig-intro} are governed by the temperature- and age-dependent response function of the spin glass. The initial magnetisation, zero in the ZFC case and $M_{FC}$ in the TRM case, does not include any component that has been attained by earlier field changes at a temperature below $T_g$; the magnetisation changes are dynamically limited and governed by the observation time that corresponds to the heating rate in the temperature dependent experiments. We also recall from these and earlier experiments, that the thermal history governs the global evolution of the spin state and the response function and that this occurs independently of any field changes within the linear response regime.\\

\begin{figure*}[h]
\begin{center}
\includegraphics[width=0.46\textwidth]{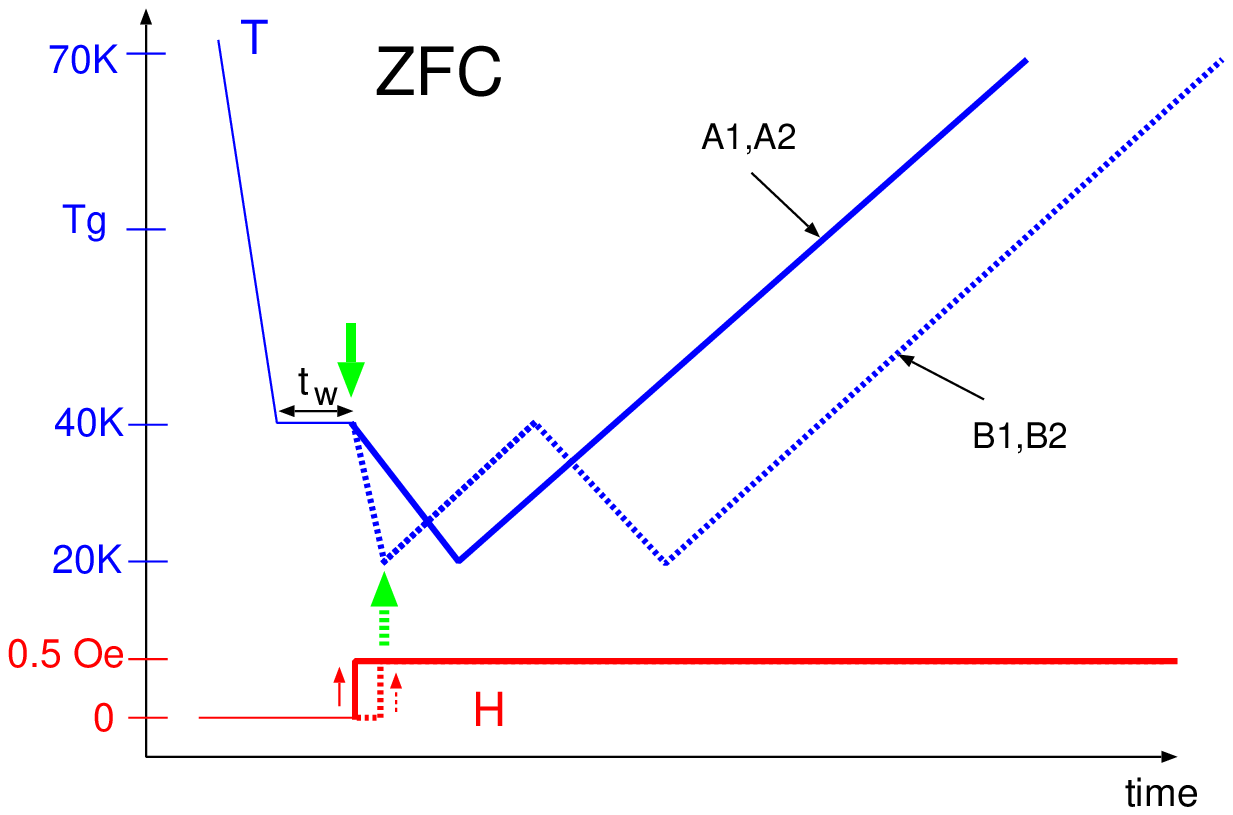}
\includegraphics[width=0.46\textwidth]{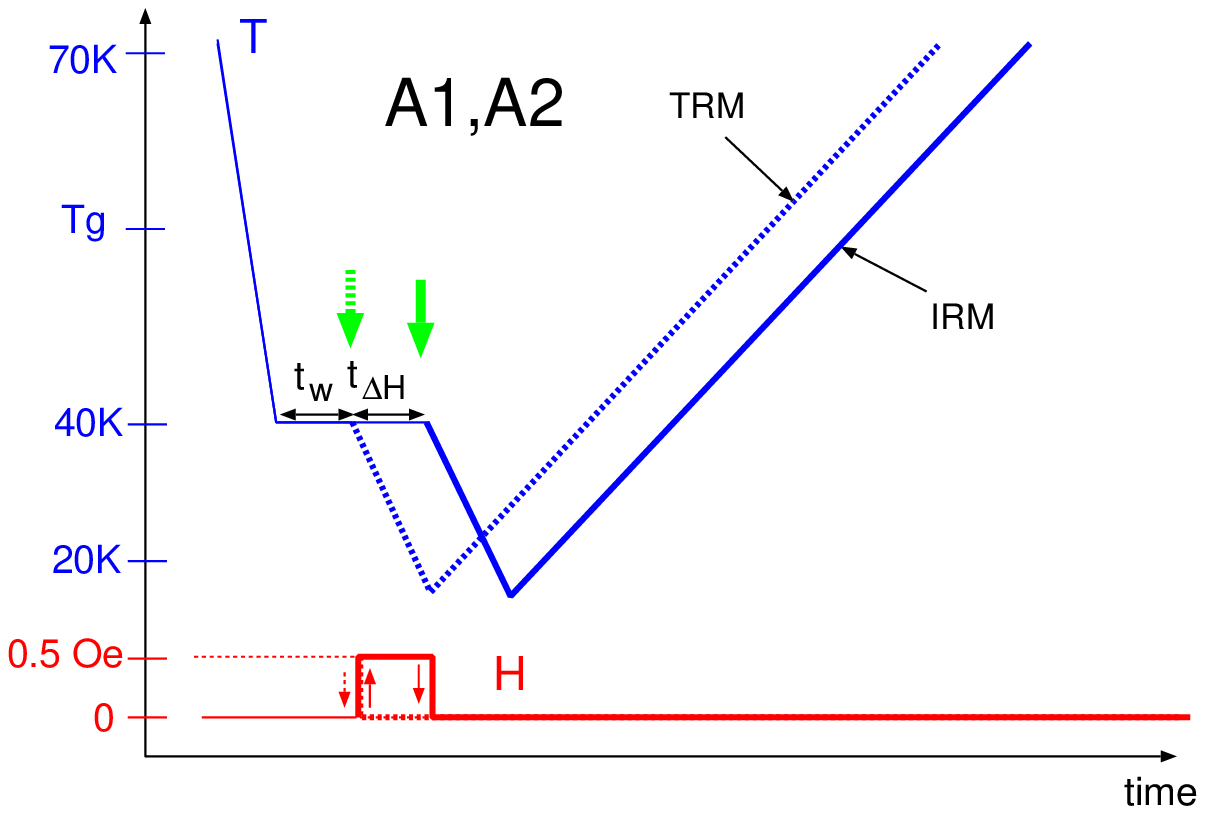}
\caption{Schematic representation of the different procedures. The arrows indicate the time a which the magnetisation starts to be recorded. Left: Variation of the temperature $T$ and magnetic field $H$ as a function of time during the procedures A1, A2 (continuous line) and B1, B2 (dotted line) in the ZFC case. In the case of A1, B1, $t_w$=0, while $t_w$=3000 s for A2, B2. Right: Idem for the procedures A1, A2 in the IRM (continuous line) and TRM (dotted line) cases.}
\label{fig-proc}
\end{center}
\end{figure*}

Let us now add in-field temperature cyclings to dc-memory experiments on the ZFC magnetisation. Table~\ref{desc} lists and Fig.~\ref{fig-proc} illustrates the different protocols that we have considered. R1 and R2 correspond to conventional ZFC experiments, without (R1) and with (R2) halt during the initial cooling, akin to the ones shown in the right panel of Fig.~\ref{fig-intro}. We will use these curves as references for the other procedures. In procedure A1, the system is cooled down below $T_g$ to 40 K (see Table~\ref{desc}). The magnetic field $H$=0.5 Oe is switched on and the magnetisation is recorded in that field on resuming the cooling down to 20 K, as well as on reheating to 70 K. Procedure A2 is identical to A1, albeit the system is kept for 3000 s at 40 K during the initial cooling (i.e. just before turning the magnetic field on) as in memory experiments. Procedure C is performed for comparison. It is similar to A2, except that the field is turned and kept on during the 3000 s long wait time at 40K. 

\begin{figure*}[t]
\begin{center}
\includegraphics[width=0.46\textwidth]{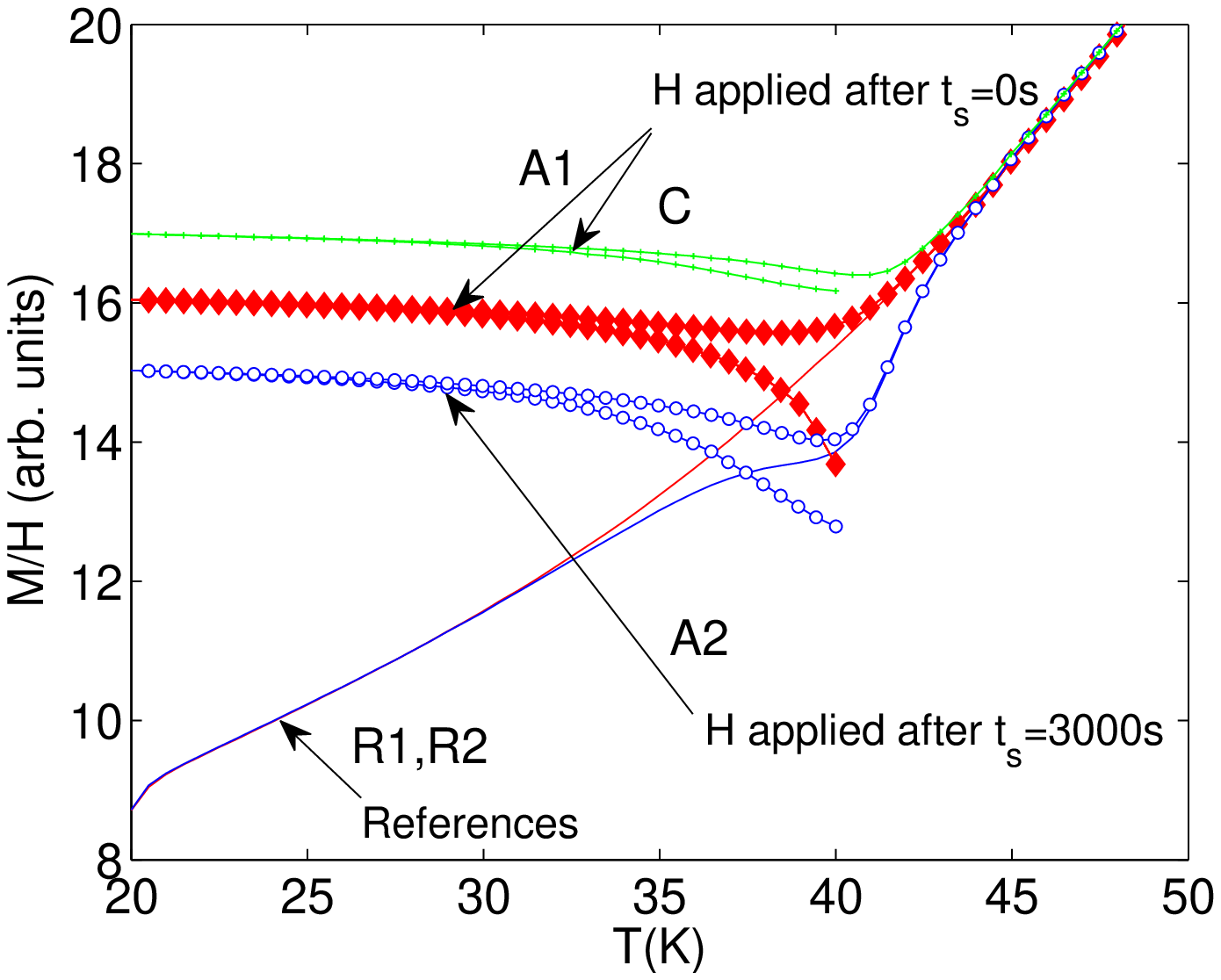}
\includegraphics[width=0.46\textwidth]{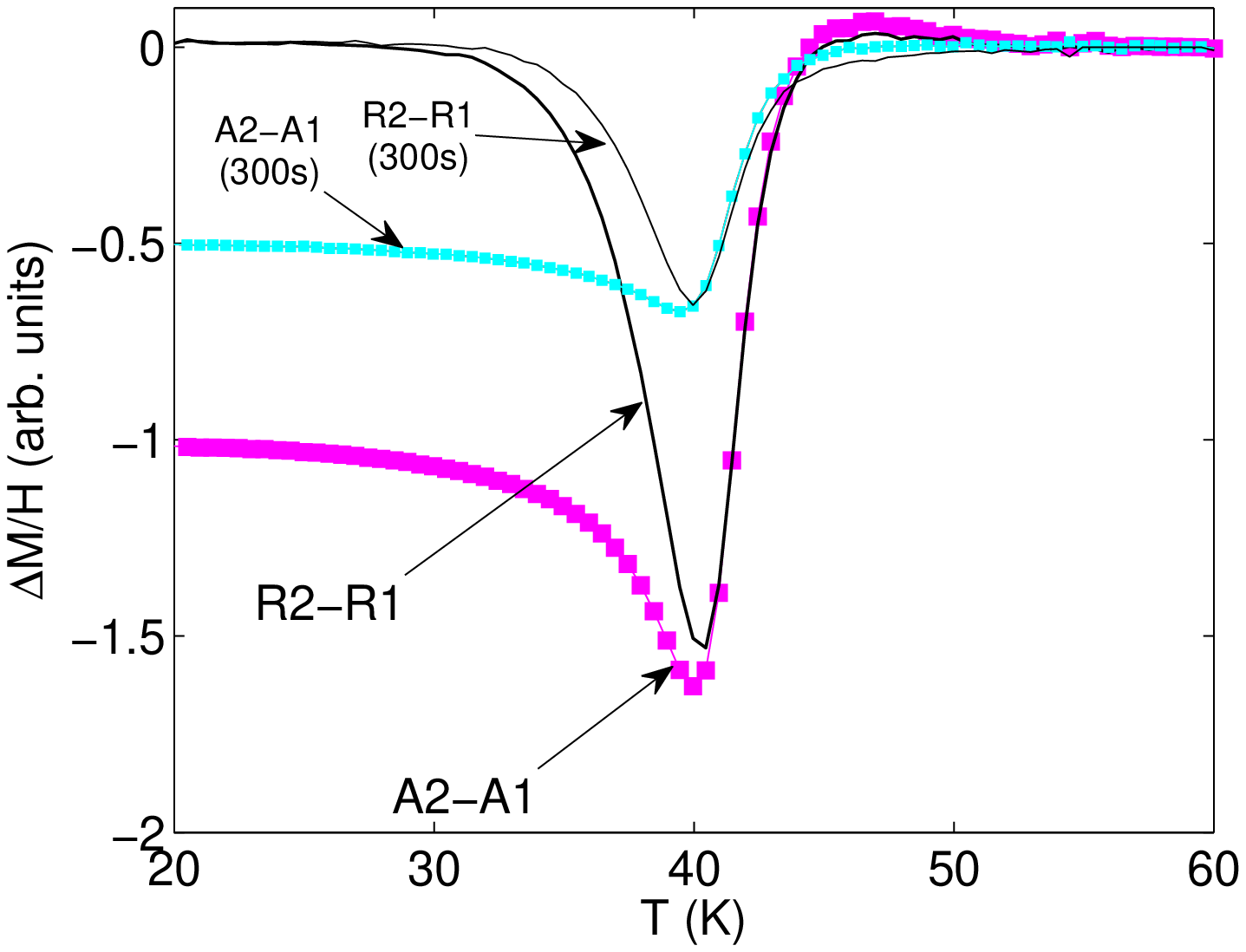}
\caption{Left: Temperature dependence of the ZFC magnetisation recorded under the A, C and R heating/cooling protocols (see table~\ref{desc} and Fig.~\ref{fig-proc}). R1 and R2 are the ZFC reference curves shown in Fig.~\ref{fig-intro}(right). Right: Difference plots of the A2-A1 curves (large symbols). The difference plots of the references R2-R1 (a conventional ``dc-memory'' plot) is added in thick continuous line. Corresponding data obtained from experiments employing a shorter wait time of 300 s are included (small symbols, thin continuous line) for comparison. Only difference plots of reheating curves are depicted.}
\label{fig-meas}
\end{center}
\end{figure*}

As seen in the left panel of Fig.~\ref{fig-meas}, the equilibration which occurred during the halt in the initial cooling (A2 or R2) is kept in memory in both types of experiments, as observed in conventional temperature cycling experiments (i.e. thermal cycling without magnetic field change) on time-dependent dc- or ac-magnetisation experiments\cite{dcising,ghost}. The final reheating curves obtained in each case eventually merge with their respective references. In the case of procedure C, the result of the large relaxation of the magnetisation akin to the one shown in the left panel of Fig.~\ref{fig-intro} can be appreciated. It can be observed in the left panel of Fig.~\ref{fig-meas} that the magnetisation curves recorded following procedures A (and C) become flat at low temperatures, with different magnetisation values. These weakly temperature-dependent magnetisation curves are reminiscent of the field-blocked field cooled \cite{petra} and TRM\cite{mamiya} magnetisation, weakly affected by the spin reorganisation on short-length scales occurring on cooling down the system. If instead we consider the difference plots of the curves recorded with and without a 3000 s halt at 40 K for the different procedures (R2-R1, A2-A1), we obtain the dc-memory curves which are shown in the right panel of  Fig.~\ref{fig-meas}. The curve labelled R2-R1 depicts a conventional dc-memory experiment, as for the inset of Fig.~\ref{fig-intro}. The other curves correspond to dc-memory experiments with in-field temperature cyclings. The frozen-in magnetisation implies that $\Delta M/H$ remains nearly constant as the temperature decrease below 40 K, outside the width of the memory dip exhibited by the reference R2-R1 curve\cite{kristian}.\\

\begin{figure*}[b]
\begin{center}
\includegraphics[width=0.46\textwidth]{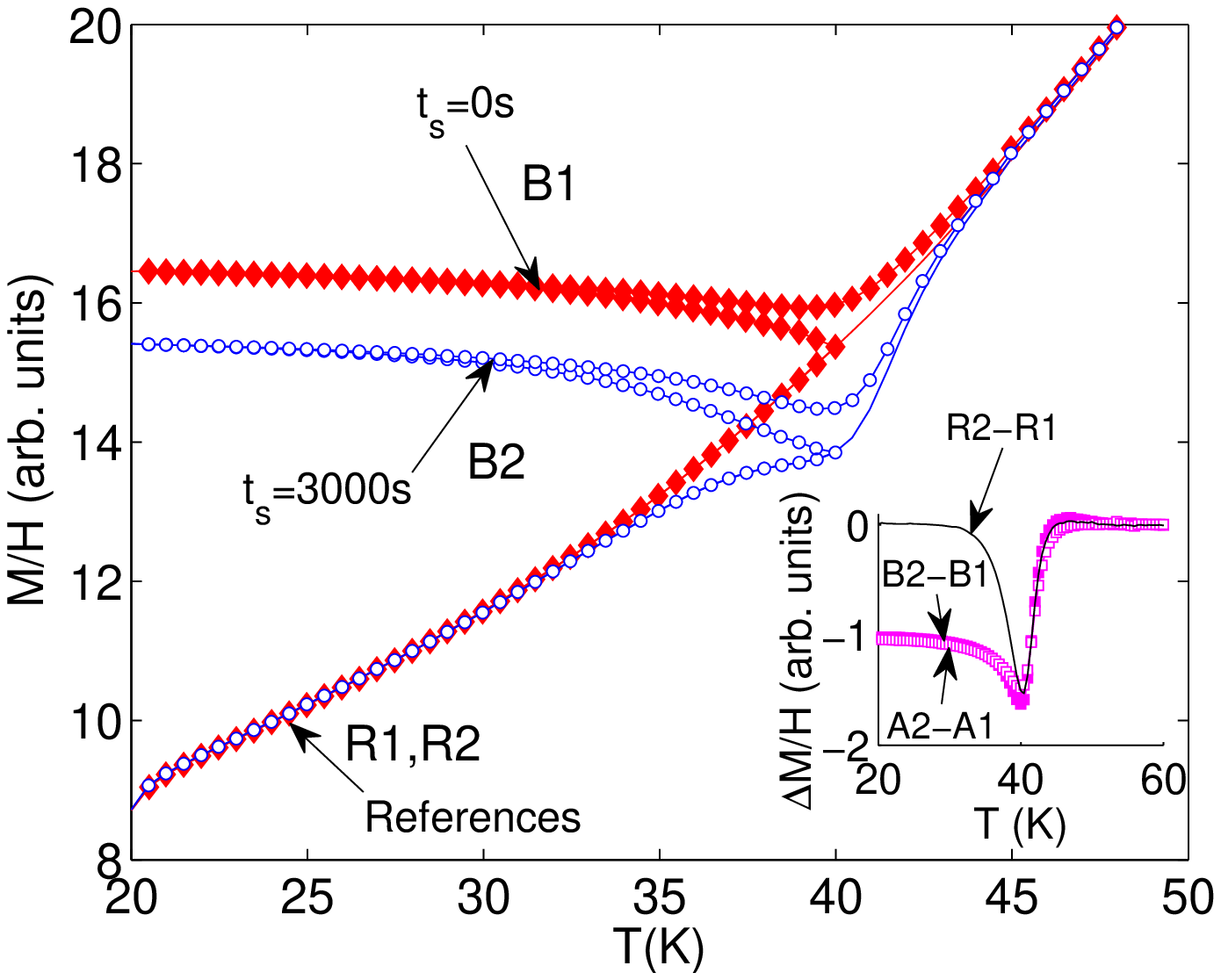}
\includegraphics[width=0.46\textwidth]{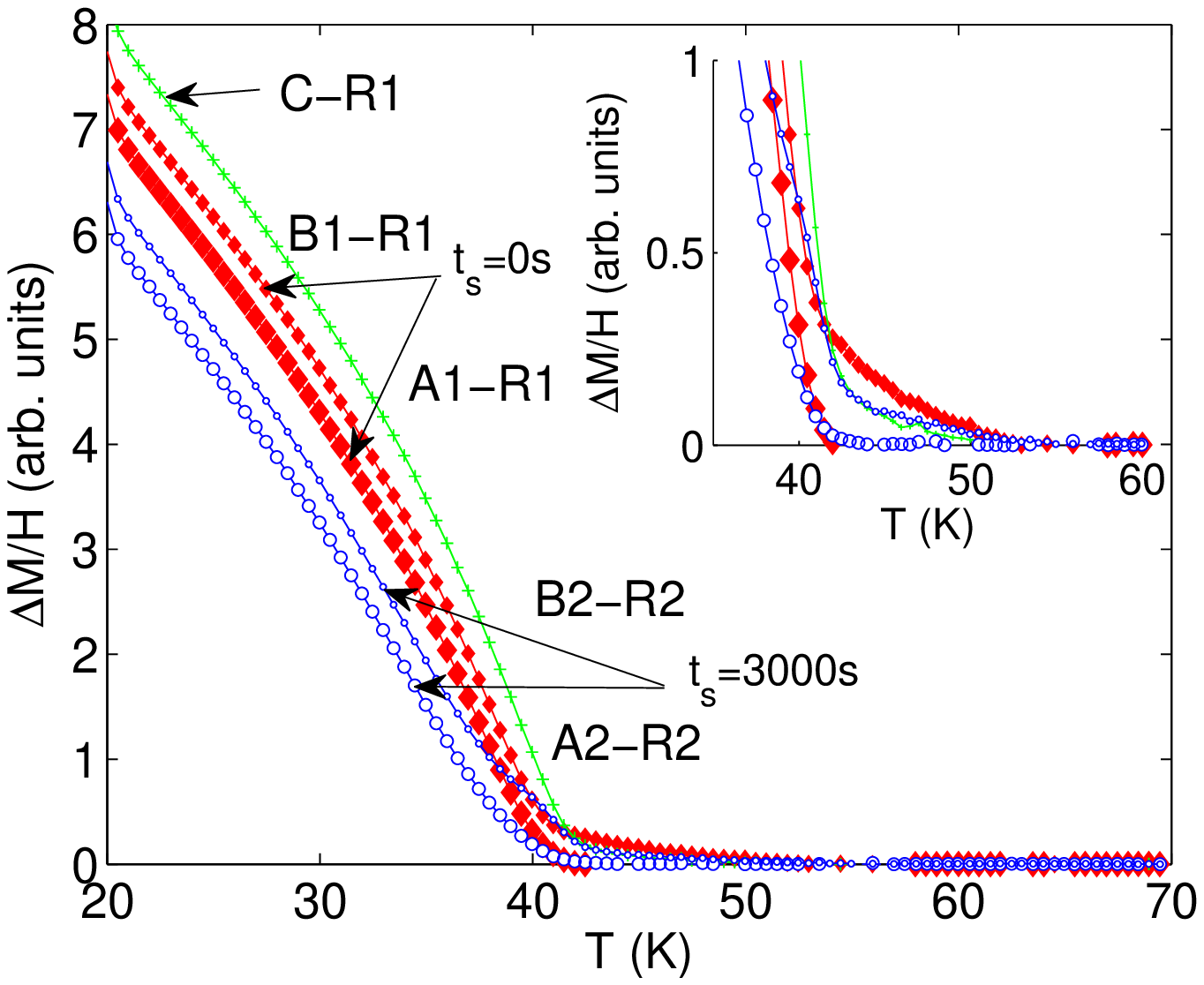}
\caption{Left: Temperature dependence of the ZFC magnetisation recorded under the B and R heating/cooling protocols (see table~\ref{desc} and Fig.~\ref{fig-proc}). The inset shows the difference plots of the B2-B1 (open symbols), A2-A1 (filled symbols), as well as R2-R1 (continuous line) curves. Right: Difference plots of the reheating A1, B1, C, A2, B2 curves with their respective reference R1 and R2. The inset shows an enlarged view of the plot in the main frame.}
\label{fig-dm}
\end{center}
\end{figure*}

In procedure A1, $M$ is recorded as soon as the temperature reaches 40 K. Let us consider another procedure, like procedure B1 (and associated B2) in which the final reheating curve is measured after first lowering the temperature to 20 K, as illustrated in the left panel of Fig.~\ref{fig-dm}. In procedure B1, the system is cooled down to 20 K, and as in a conventional ZFC measurement (R1), the magnetic field $H$=0.5 Oe is switched on and the magnetisation is recorded in that field on reheating. In this case however the reheating stops at 40 K. While still recording the magnetisation in $H$=0.5 Oe, the system is cooled again down to 20 K, and reheated to 70 K. Procedure B2 is identical to B1, albeit the system is kept for 3000 s at 40 K during the initial cooling as in memory experiments.

As for procedures A1 and A2, the final reheating curves obtained in each case eventually merge with their respective references. The magnetisation curves become flat at low temperatures, with different magnetisation values (with $M_{\rm C} > M_{\rm B1} > M_{\rm A1} > M_{\rm B2} > M_{\rm A2}$, see also the right panel of Fig.~\ref{fig-dm}). As seen in the right panel of Fig.~\ref{fig-dm} which shows the difference plots of the reheating curves obtained for the five procedures, regardless of the halt at 40 K during the initial cooling, the reheating curves measured on the system using protocols A1 and A2 merge earlier with the reference than those measured using protocols B1 and B2. In the latter case, a measurable $\Delta M/H$ is observed way above 40 K. This subtle difference of the essentially ZFC magnetisation that is recorded well above the temperature 40 K (below which the thermal and field history differs in protocols A and B) is mainly associated with the increase of the magnetisation that occurs around 40 K in protocol B before the second cooling to 20 K.\\

\begin{figure*}[t]
\begin{center}
\includegraphics[width=0.46\textwidth]{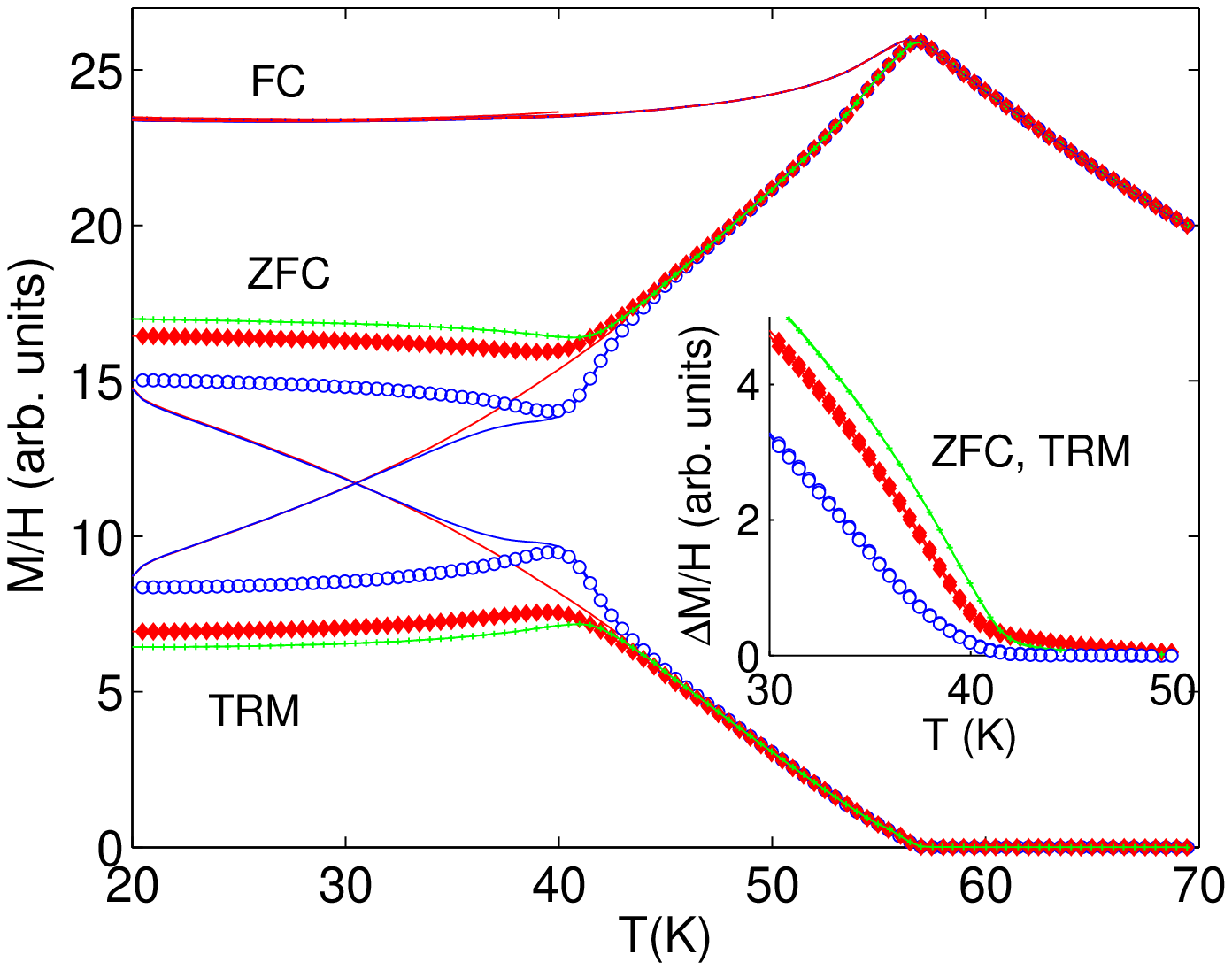}
\includegraphics[width=0.46\textwidth]{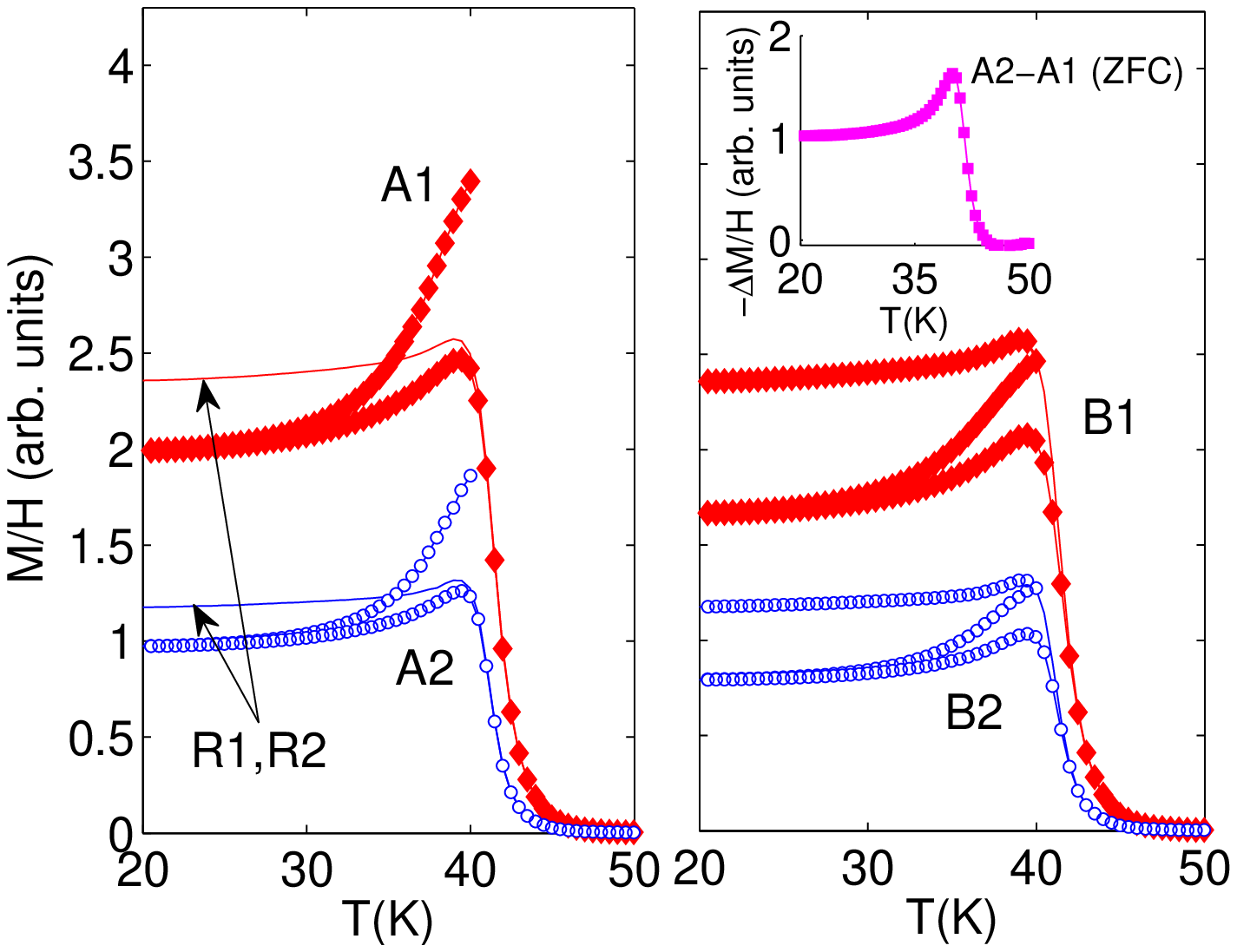}
\caption{Left: Temperature dependence of the ZFC, FC, and TRM magnetisation recorded under the same heating/cooling protocols. For clarity no symbols are employed to plot the $M_{\rm FC}$ data, and only reheating data for procedures A2 (open symbols), C (pluses), and B1 (filled symbols), and reference curves (continuous lines) are shown. The inset shows the corresponding difference plots (A2-R1, C-R1, and B2-R2) for the ZFC and TRM (plotted as $-M_{\rm TRM}$) data, which are virtually identical in all three cases. Right: Temperature dependence of the IRM magnetisation recorded under the same heating/cooling protocols. In the reference IRM measurements, a magnetic field of $H$=0.5 Oe is applied for 3000 s at 40 K and removed, and $M$ is recorded (in $H$=0) on reheating to 70 K after resuming the cooling to 20 K (also in $H$=0). The A1, A2, B1 and B2 procedures are performed akin to the ZFC case, e.g. in the case of A1 by starting to measure $M$ in $H$=0 after the field cycling at 40 K, on cooling to 20 K and subsequent reheating to 70 K.The inset shows the difference plot A2-A1 obtained for ZFC experiments for comparison (from Fig.~\ref{fig-dm}, plotted as $-\Delta M$)}
\label{fig-more}
\end{center}
\end{figure*}

Illustrated in the left panel of Fig.~\ref{fig-more}, are  ZFC, FC and TRM curves recorded using the same thermal protocols (R, A, B and C) as described above for the ZFC case. The TRM behaviour corresponds nicely with the ZFC behaviour, whereas  there is  little effect of wait times and temperature cyclings on the FC magnetisation. As seen in the inset of the left panel of Fig.~\ref{fig-more}, we can see that the difference curves obey as well the linear relation mentioned earlier, with $\Delta M_{ZFC} \sim - \Delta M_{TRM}$.

In the right panel of Fig.~\ref{fig-more} results of the IRM after similar thermal protocols R, A, and B are shown. IRM refers to isothermal remanent magnetisation\cite{dcagmn} and is measured in $H$=0 after cooling the system also in $H$=0; a halt is performed at constant temperature for a time $t_{\Delta H}$ during which a magnetic field $\Delta H$ is applied (See Fig.~\ref{fig-proc}). The IRM experiments are interesting, because in that case the magnetic field is applied only to imprint the IRM during the halt (and turned off elsewhere). The behaviour of the IRM curves looks at a glance similar to those of the A2-A1 and B2-B1 curves of the right hand panel of Fig.~\ref{fig-dm}, as illustrated in the inset of Fig.~\ref{fig-more}. This may be anticipated; it is however worth to note that the thermal history differs in the IRM case, as there is always an additional 3000 s stay at 40 K during which the field is applied. The magnetisation that is attained during the field application is frozen in on cooling. This magnetisation has a fundamentally different nature than that of the dynamically controlled magnetisation that occurs with time at constant temperature after a field change or on heating after a field change. A frozen-in magnetisation state is also attained when the sample is cooled and reheated from 40 to 20 K in protocol A and B. The weak temperature-dependence of the FC, ZFC, TRM and IRM curves between 20 and 35 K mimics frozen-in magnetisation states.  The frozen-in magnetisation rapidly fades away when the temperature is increased above the temperature where it is attained.

Looking again at the right panel of Fig.~\ref{fig-more} one notices a sharp upward temperature dependence on approaching 40 K on all the reheating curves. IRM curves with similar features were observed for another canonical Ag(Mn) spin glass\cite{dcagmn}. The spin configuration that was originally imprinted by the magnetic field at 40 K brings forth an excess magnetisation in the direction of the applied field. This excess magnetisation is slowly drained by random spin fluctuations without preferred direction as the temperature is shifted away from 40 K. However, when coming back towards 40 K the memory of the spin configuration attained during the in-field stop at this temperature is recovered and this includes the unidirectional excess moment, of which only a minor part has been lost by the random spin fluctuations during the temperature cycling to lower temperatures. Above 40 K the excess moment rapidly fades away as the memory of the low temperature configuration is washed out.\\

\textbf{Conclusion.} -  The temperature dependence of the magnetisation of a Cu(Mn) spin glass on its thermal and magnetic field history has been studied using various protocols. The results show that rejuvenation does not affect the ``frozen-in''  magnetisation. On the other hand, the thermal history on cooling, the heating rate and the wait time at constant temperatures govern the response function, that is reflected in the ordinary ZFC (or TRM) magnetisation response.

A major finding is that the spin configuration that is imprinted during elongated stop during cooling, if it is accompanied by a field-induced magnetisation, also includes a unidirectional excess magnetisation that is recovered on returning to the ageing temperature. This occurs in spite of the fact that substantial parts of that magnetisation appears to be lost through spin fluctuations at lower temperatures. Even when rapid relaxation due to slow cooling occurs, this unidirectional excess magnetisation is almost perfectly recovered when the sample returns to the temperature where it was aged (see right panel of Fig.~\ref{fig-more}).\\

\acknowledgments
We thank the Swedish Research Council (VR) and the G\"oran Gustafsson Foundation for financial support.

\begin{table*}[h]
\caption{Description of the different cooling and measurement protocols employed in the study. Procedures R2, A2 and B2 are identical to R1, A1, and B1 respectively, adding a halt at 40 K for 3000 s during the initial cooling down to 20 K. In procedure C, the halt is performed with $H$ $\neq$ 0.}
\label{desc}
\begin{center}
\begin{tabular}{c|c|l|l}
& Initial cool down & & Measure $M(T)$ in $H$=0.5 Oe \\
Procedure & in $H$=0 to $T$= & Wait at 40 K & with $T$ varying as: \\
\hline
R1 & 20 K & 0 s & 20 K $\rightarrow$ 70 K \\
R2 & 20 K & 3000 s ($H$ = 0) & 20 K $\rightarrow$ 70 K \\
\hline
A1 & 40 K & 0 s & 40 K $\rightarrow$ 20 K $\rightarrow$ 70 K \\
A2 & 40 K & 3000 s ($H$ = 0) & 40 K $\rightarrow$ 20 K $\rightarrow$ 70 K \\

C & 40 K & 3000 s ($H$ = 0.5 Oe) & 40 K $\rightarrow$ 20 K $\rightarrow$ 70 K \\
\hline
B1 & 20 K & 0 s & 20 K $\rightarrow$ 40 K $\rightarrow$ 20 K $\rightarrow$ 70 K\\
B2 & 20 K & 3000 s ($H$ = 0) & 20 K $\rightarrow$ 40 K $\rightarrow$ 20 K $\rightarrow$ 70 K

\end{tabular}
\end{center}
\end{table*}

\end{document}